\documentclass[traditabstrac]{aa}  

\usepackage{graphicx}
\usepackage{natbib}
\usepackage{txfonts}
\usepackage{color}

\newcommand{\Hermes} {\textsc{Hermes}\xspace}
\newcommand{\hermes} {\textsc{Hermes}\xspace}

\newcommand{\sindex}{$\mathcal{S}$-index\xspace}
\newcommand{\Sindex}{$\mathcal{S}$-index\xspace}
\newcommand{\Ssymbol}{$\mathcal{S}$\xspace}

\usepackage[switch]{lineno} % line number
\begin{document}

   %\title{Magnetic activity variability  in the young solar analog KIC\,10644253}
   \title{Magnetic variability in the young solar analog KIC\,10644253}
   \subtitle{Observations from the \emph{Kepler} satellite and the \Hermes spectrograph}
   
   \author{D. Salabert  \inst{1,2}
         \and
        C. R\'egulo \inst{3,4}
         \and
         R.~A. Garc\'ia \inst{1}
        \and
         P.~G. Beck \inst{1}
         \and 
         J. Ballot \inst{5,6}
          \and
         O.~L. Creevey \inst{2}
         \and
         F. P\'erez Hern\'andez \inst{3,4}    
           \and 
         J.-D. do Nascimento Jr. \inst{7,8}
          \and   
          E. Corsaro \inst{1,3,4}
         \and 
         R. Egeland \inst{9,10}
	 \and
	 S. Mathur \inst{11}
	 \and
          T.~S. Metcalfe \inst{11}
	 \and
	 L. Bigot \inst{2}
	 \and
         T. Cellier \inst{1}
         \and	     
         P.~L. Pall\'e\inst{3,4}
         }

%	    		M.~F. Andersen\inst{5}
%	\and
	%	\and
%	A. Trivi\~no Hage\inst{6,7} 
         
%     }

   \institute{Laboratoire AIM, CEA/DRF-CNRS, Universit\'e Paris 7 Diderot, IRFU/SAp, Centre de Saclay, 91191 Gif-sur-Yvette, France\\
              \email{david.salabert@cea.fr}
              \and
	Laboratoire Lagrange, UMR7293, Universit\'e de la C\^ote d'Azur, CNRS, Observatoire de la C\^ote d'Azur, Nice, France
            	\and
    	Instituto de Astrof\'isica de Canarias,  E-38200 La Laguna, Tenerife, Spain
         \and
      	Departamento de Astrof\'isica, Universidad de La Laguna, E-38205 La Laguna, Tenerife, Spain
	\and
	CNRS, Institut de Recherche en Astrophysique et Plan\'etologie, 14 avenue Edouard Belin, 31400 Toulouse, France
	\and
	Universit\'e de Toulouse, UPS-OMP, IRAP 31400, Toulouse, France
		\and
       Universidade Federal do Rio Grande do Norte, UFRN, Dep. de F\'{\i}sica, DFTE, CP1641, 59072-970, Natal, RN, Brazil 
       \and	
	Harvard-Smithsonian Center for Astrophysics, Cambridge, Massachusetts 02138, USA
	\and 
	High Altitude Observatory, National Center for Atmospheric Research, PO Box 3000, Boulder, CO 80307-3000, USA
	\and
	Department of Physics, Montana State University, Bozeman, MT 59717-3840, USA
	\and
	Space Science Institute, 4750 Walnut street Suite\#205, Boulder, CO 80301, USA}
	
   \date{Received XX; accepted XX}

% \abstract{}{}{}{}{} 
% 5 {} token are mandatory
 
  \abstract
    {The continuous photometric observations collected by the {\it Kepler} satellite over 4 years provide a whelm of data with an unequalled quantity and quality for the study of stellar evolution of more than 200\,000 stars. Moreover, the length of the dataset provide a unique source of information to detect magnetic activity and associated temporal variability in the acoustic oscillations. In this regards, the {\it Kepler} mission was awaited with great expectation. The search for the signature of magnetic activity variability in solar-like pulsations still remained unfruitful more than 2 years after the end of the nominal mission. Here, however, we report the discovery of temporal variability in the low-degree acoustic frequencies of the young (1\,Gyr-old) solar analog KIC\,10644253 with a modulation of about 1.5 years with significant temporal variations along the duration of the {\it Kepler} observations. The variations are in agreement with the derived  photometric activity. The frequency shifts extracted for KIC\,10644253 are shown to result from the same physical mechanisms involved in the inner sub-surface layers as in the Sun. In parallel, a detailed spectroscopic analysis of KIC\,10644253 is performed based on complementary ground-based, high-resolution observations collected by the \hermes instrument mounted on the \textsc{mercator} telescope. Its lithium abundance and chromospheric activity \Sindex confirm that KIC\,10644253 is a young and more active star than the Sun. }
  
               
   \titlerunning{Magnetic variability in the young solar analog KIC\,10644253}

   \maketitle
%
%________________________________________________________________

%------------------------------------------------------------------------------------------------------------------------------
\section{Introduction}
Photometric observations of the solar-like oscillator G0V KIC\,10644253 (BD+47\,2683, $V=9.26$, $[\text{Fe/H}] = 0.12$\,dex, and see Table\,\ref{tab:properties}) were collected continuously over 4 years between May 2009 and May 2013 by the {\it Kepler} satellite \citep{borucki10}. At first, its stellar properties were derived from grid-based modeling based on global seismic parameters combined with complementary photometry \citep{pins12} and spectroscopy \citep{bruntt12} as part of more than 500 solar-like {\it Kepler} targets \citep{chaplin14}. Because its effective temperature,  above 6000\,K \citep{bruntt12}, KIC\,10644253 was part of the sample of 22 hot stars selected by \citet{mathur14a} where they reported a modulation in the temporal variation of its photometric magnetic activity.
 In the mean time, more precise stellar properties based on the fitted individual oscillation frequencies from \citet{appourchaux12} and derived using the Asteroseismic Modelling Portal \citep[AMP;][]{metcalfe14} substantially revised downwards its mass by more than 20\% to $1.13\pm0.05\,M_{\sun}$ with an estimated age of  $1.07\pm0.25$\,Gyr. Making the best use of the asteroseismic observations, the stellar parameters of KIC\,10644253 thus correspond to a solar analog, the youngest solar-like pulsating star observed by {\it Kepler} with a rotation period of $10.91\pm0.87$\,days \citep{garcia14a}. Furthermore, the unequalled length of the {\it Kepler} observations provide as well a great opportunity to study stellar temporal variations by means of different observables, such as the surface magnetic activity \citep[e.g.,][]{frasca11,frohlich12,bonanno14a}.
 As the proxy for photometric magnetic activity suggests that KIC 10644253
is among the most active stars of the solar-like oscillators
observed by {\it Kepler} \citep{garcia14a}, and due to its very solar-like
fundamental stellar properties,
KIC 10644253 is an excellent candidate to investigate the magnetic activity of a young Sun
with asteroseismic data.

Indeed, it is well established that in the case of the Sun, acoustic (p) oscillation frequencies are sensitive to changes in the surface activity and vary in correlation with activity proxies such as the sunspot number or the  10.7 cm radio emission along the 11-year solar cycle. Such temporal variations were quickly reported with the first helioseisimic observations by \citet{wood85}, and confirmed shortly after by \citet{fossat87} and \citet{libbrecht90}. As longer and higher-quality helioseismic observations became available, the temporal variability of the p-mode frequencies  were studied in greater detail. These frequency shifts were actually observed to be frequency dependent, the shifts being larger at higher frequencies, and to be angular-degree ($l$) dependent, i.e., mode-inertia dependent, although the  $l$ dependence is  small for low-degree modes  \citep[see, e.g.,][]{gelly02,howe02,chano04,salabert04}. 
Moreover, \citet{howe02} showed that the temporal and latitudinal distribution of the frequency shifts is correlated with the spatial distribution of the surface magnetic field. The frequency shifts are explained to arise from indirect magnetic effects on the inner structural changes \citep{kuhn88,dziem05} of the Sun along the solar cycle. Today, solar frequencies are the only activity proxy which can reveal inferences on sub-surface changes with solar activity not detectable at the surface by standard global proxies \citep[e.g.,][]{salabert09,salabert15,basu12}. Furthermore, a faster temporal modulation of the oscillation frequency variations, with a period of about 2 years, was observed to coexist in the Sun \citep{fletcher10}, which is believed to have an origin located deeper inside the convective zone. Except the Sun, temporal variations of p-mode frequencies related to magnetic activity using asteroseismic techniques are so far observed in only one star: the F5V star HD\,49933  observed by the space-based Convection, Rotation, and planetary Transits \citep[CoRoT,][]{baglin06} satellite. Indeed, \citet{garcia10} reported for the first time that the acoustic oscillation parameters were varying in a comparable manner as in the Sun, and in correlation with the photometric measurements. The frequency dependence of the activity-frequency relationship was also reported to be similar as the one observed in the Sun, but with larger shifts in accordance with a larger activity index \citep{salabert11}. Such asteroseismic inferences of the magnetic activity  provide invaluable information when studying the physical phenomena involved in the dynamo processes in the interior of stars and in this regards the 4-year long continuous observations collected by the {\it Kepler} mission were awaited with great expectation for detecting activity cycles. The CoRoT target HD\,49933 remains nonetheless the only object so far with the Sun for which changes in the inner sub-surface layers related to magnetic activity were observed using asteroseismology.

Nevertheless, a large sample of stellar magnetic cycles covering a range between 2.5 and 25~years were reported by several authors \citep[e.g.,][]{Wilson78,Baliunas85,Baliunas95,Hall07} from spectroscopic measurements in the \ion{Ca}{ii} H and K emission lines at the Mount Wilson Observatory (MWO). Moreover, several authors suggested that the periods of the activity cycles increase proportional to the stellar rotational periods along two distinct paths in main-sequence stars: the active and the inactive stars \citep{Saar99,BV07}. 
Furthermore, 9 stars with dual cycles were identified within the MWO survey \citep{Baliunas95}, 6 of which made it into the refined sample of \citet{Saar99}, along with one newly reported dual-cycle star. Another 15 stars with multiple cycles were reported by \citet{olah09} using photomometric and \ion{Ca}{ii} emission observations. Recently, \citet{metcalfe10} observed short-period variations of 1.6\,years on the exoplanet host star $\iota$~Horologi, and dual cycles, respectively of 2.95 and 12.7\,years, were found on $\epsilon$~Eridani, an exoplanet host star as well, by \citet{metcalfe13} with the high cadence SMARTS HK observations.
Recently, \cite{egeland15} observed  Sun-like magnetic temporal variations in the G1.5V star HD\,30495, a 1\,Gyr-old solar analog with a rotation period of $\sim$\,11\,days. They showed that this star has short-period variations of $\sim$\,1.7\,years and a long modulation of $\sim$\,12\,years.

 %%%%%%%%%%%%%%%%%%%%%%%%%%%%%%%
\begin{table}[t]
\caption{KIC\,10644253 stellar properties.}
\centering         
\begin{tabular}{lll}        
\hline\hline
Property &  Value & Reference \\
   \hline
   Spectral type & G0 V & (1)\\
   $V$ & $9.26\pm0.02$ & (2)\\
   $(B-V)$ & $0.59\pm0.03$ & (2)\\
\hline
Mass ($M_{\sun}$) & $1.13\pm0.05$ & (3) \\
Radius $R$ ($R_{\sun}$)& $1.108\pm0.016$ & (3) \\
Age (Gyr)& $1.07\pm0.25$ & (3) \\
$R_\mathrm{\textsc{bcz}}/R$ & $0.77\pm0.03$& (3)\\
\hline
 $\upsilon \sin i_\mathrm{spectro}$ (km s$^{-1}$) & $3.8\pm0.6$  & (4) \\
 $T_\mathrm{eff}$ (K) &  $6030\pm60$ & (4)\\
log $g$ (dex) & $4.40\pm0.03$& (4)\\
$[\text{Fe/H}]$ (dex) & $0.12\pm0.06$ & (4)\\

$\upsilon \sin i_\mathrm{astero}$ (km s$^{-1}$) & 0.62\,$\pm$\,0.81  & (5) \\
$\nu_\mathrm{max}$ ($\mu$Hz) & $2819\pm131$ & (6)\\	
$\Delta\nu$ ($\mu$Hz) & $123.6\pm2.7$ & (6)\\	
$P_\text{rot}$ (days) & $10.91\pm0.87$ & (7)\\
\hline
$i_\mathrm{spectro}$ ($^\circ$) & 48$^{+11}_{-9}$ & (3+4+7)\\
$i_\mathrm{astero}$ ($^\circ$) & 7$^{+9}_{-9}$ & (3+5+7)\\

 \hline
\end{tabular}
 \tablebib{(1) \citet{molenda13}; (2) \citet{hog00}; (3) \citet{metcalfe14}; (4) \citet{bruntt12}; (5) \citet{doyle14}; (6) \citet{chaplin14}; (7) \citet{garcia14a}; (3+4+7) Derived from \citet{metcalfe14}, \citet{bruntt12}, and \citet{garcia14a}; (3+5+7) Derived from  \citet{metcalfe14}, \citet{doyle14}, and \citet{garcia14a}.}
\label{tab:properties}
\end{table}%
%%%%%%%%%%%%%%%%%%%%%%%%%%%%%%%

Another way to detect magnetic cycles is to directly measure the evolution of the photospheric magnetic field through Zeeman effect. 
Activity cycles were found this way in late-type dwarf stars that exhibit periodic polarity reversal of their magnetic field as observed in the Sun during its 11-year cycle. For example, the F-type star $\tau$\,Boo and HD\,78366, respectively rotating in 3.3 and 11.4\,days, magnetic cycles of about 2 years \citep{fares09} and shorter that 3 years \citep{morg11} were respectively observed. However, these periods can be different to the variations measured in chromospheric proxies: for instance, \citet{Baliunas95} reported two periods of 5.9 and 12.2\,years in the chromospheric emission of HD\,78366. It is then important to keep in mind that stellar magnetic activity cycles can be multi-periodic and that different varieties of observations provide complementary constraints on the underlying dynamo processes.

In that context, we investigate in this work the magnetic activity of the young solar analog KIC\,10644253 by means of the photometric and high-resolution spectroscopic observations collected respectively by the {\it Kepler} satellite and the \Hermes instrument \citep{raskin11,raskin11phd}. 
In Section\,2, we describe the {\it Kepler} and \Hermes data used in this paper. The methods applied to derive the photometric activity and the low-degree seismic acoustic frequencies and how the frequency shifts are estimated are described in Section\,3. The results are presented and discussed in Section\,4. 
In Section\,5, the magnetic activity and age of KIC\,10644253 are discussed in the light of complementary spectroscopic \Hermes observations. In Section\,6, we provide interpretations on the  inner structural changes of KIC\,10644253 with activity and compare to the Sun, while a summary of the results is given in Section\,7.  
 
%%%%%%%%%%%%%%%%%%%%%%%%%%%%%%%
\begin{table*}[t!]
\caption{Journal of \Hermes observations for KIC\,10644253.}
\centering
\begin{tabular}{ccccccc}
\hline
\hline
\# spectrum & HJD & Exposure Time & Radial Velocity    & \Ssymbol & SNR(Ca)\\
 & (day) & (s) & (km s$^{-1}$) &  & \\
\hline
1 & 2457087.68434 & 1600 & $-19.00\pm0.04$ & 	0.207 & 27.9  \\
2$^\dagger$ & 2457087.77404 & 1700 & $-19.00\pm0.04$ & 	0.221 & 38.2  \\
3$^\dagger$ & 2457088.71177 & 1800 & $-19.00\pm0.04$ & 	0.213 & 16.5  \\
4$^\dagger$ & 2457088.73319 & 1800 & $-19.00\pm0.04$ & 	0.243 & 21.5  \\
5$^\dagger$ & 2457090.73719 & 1600 & $-19.02\pm0.04$ & 	0.248 & 34.4  \\
6 & 2457090.75629 & 1600 & $-19.02\pm0.04$ & 	0.221 & 33.7  \\
7 & 2457130.72239 & 1800 & $-18.92\pm0.04$ & 	0.209 & 38.9  \\
8 & 2457163.60000 & 1600 & $-19.01\pm0.04$ &   0.219 & 30.3  & \\
9 & 2457189.62647 & 1600 & $-18.96\pm0.04$ & 	0.204 & 38.2  & \\
10 & 2457215.45761 & 1600 & $-18.95\pm0.04$ & 0.214  & 31.6  & \\
11 & 2457241.48984 & 1000 & $-19.07\pm0.04$ & 	0.207 & 25.3  & \\
12 & 2457267.39006 & 1000 & $-19.07\pm0.04$ & 	0.226 & 24.2  & \\
\hline
\end{tabular}
\tablefoot{The Heliocentric Julian Date (HJD) is given for the midpoint of each exposure. The \Sindex values obtained for each  individual spectrum are also reported. The SNR(Ca) in the range of the UV was calculated from the instrumental flux, corrected for the gain of the CCD at the centre of the 90$^\mathrm{th}$ echelle order ($\sim$\,400\,nm). The spectra marked with $^\dagger$ were obtained during poor weather conditions (source: from the database of the Liverpool telescope at La Palma, Canary Islands, Spain).}
\label{tab:rv}
\end{table*}
%%%%%%%%%%%%%%%%%%%%%%%%%%%%%%%

%------------------------------------------------------------------------------------------------------------------------------
\section{Observations}
\label{sec:obs}
Both the long-cadence (LC) and short-cadence (SC) observations\footnote{The Data Release 23 (DR23) short-cadence data were used in this analysis, and thus are not suffering from the calibration error recently discovered by the {\it Kepler}/NASA team which only affects DR24.} of KIC\,10644253 collected by the {\it  Kepler} satellite were used in this analysis.
The LC data were analyzed over almost the entire duration of the mission from 2009 June 20 (Quarter\,2, hereafter Q2) to 2013 May 11 (Q17) (i.e., a total of about 1411 days). The two first quarters Q0 ($\sim$\,10 days) and Q1 ($\sim$\,33 days) were dropped due to calibration issues as it is performed quarter by quarters (see below). The light curve was corrected for instrumental problems using the \emph{Kepler} Asteroseismic Data Analysis and Calibration Software \citep[KADACS,][]{garcia11}. The time series was then high-pass filtered with a triangular smooth function for periods longer than 55 days in a quarter by quarter basis.  To reduce the effects of the regular gaps in the following analysis, all the gaps with a size smaller than 20 days were interpolated using an in-painting technique \citep[for more details, see][]{garcia14b,pires15}. With a sampling rate of 29.4244\,min, the signature of the photospheric magnetic activity can be inferred by studying the temporal fluctuations of the light curve.

The SC data from Q5 (2010 March 20) to Q17 were used for the seismic analysis. Indeed, although 33 days of observations were collected during Q1, KIC\,10644253 was not observed in SC during the three consecutive quarters Q2, Q3, and Q4. The SC sampling rate of 58.84876\,sec (Nyquist frequency of $\sim$\,8.50\,mHz) is high enough to measure the low-degree p-mode oscillations centered around $\nu_\mathrm{max}=2819\pm131\mu$Hz \citep{chaplin14}. As for the LC data, the SC data were corrected using the KADACS pipeline, high-pass filtered with a triangular smooth function for periods longer than 2 days, and the gaps of sizes smaller than 3 days interpolated using the in-painting technique. \\

In addition to {\it Kepler} photometry, high-resolution spectroscopic observations of the solar analog KIC\,10644253 were obtained in 2015 with the \Hermes spectrograph, mounted to the 1.2\,m \textsc{Mercator} telescope on La Palma (Canary Islands, Spain). Over a period of 180 days, 12 spectra were taken. The journal of observations is given in Table\,\ref{tab:rv}. The data reduction of the raw data was performed with the instrument-specific pipeline \citep{raskin11}, providing final data products which were rebinned to a constant resolving power of R\,=\,85\,000 over the wavelength range in between 390 and 900\,nm. The radial velocity for each individual spectrum was determined from a cross correlation with the G2-mask in the \Hermes pipeline toolbox. The radial velocity remains stable at $-19.00\pm0.04$\,km s$^{-1}$, which excludes the presence of a stellar companion given the $1\sigma$ night-to-night stability of 70\,m s$^{-1}$ of the \Hermes spectrograph. Before combining the individual spectra through the median to a high signal-to-noise-ratio (SNR) spectrum, the spectra were normalized, utilizing a solar spectrum and shifted to the rest wavelength. The compiled median spectrum with an accumulated integration time of 5.2 hours exhibits a SNR\,$\sim$\,280 in the wavelength range corresponding to Str\"omgren $y$. Details about the post processing of the \Hermes data can be found in \citet{Beck15}.

% -------------------
\begin{figure*}[tbp]
\begin{center} 
\includegraphics[width=0.98\textwidth,angle=0]{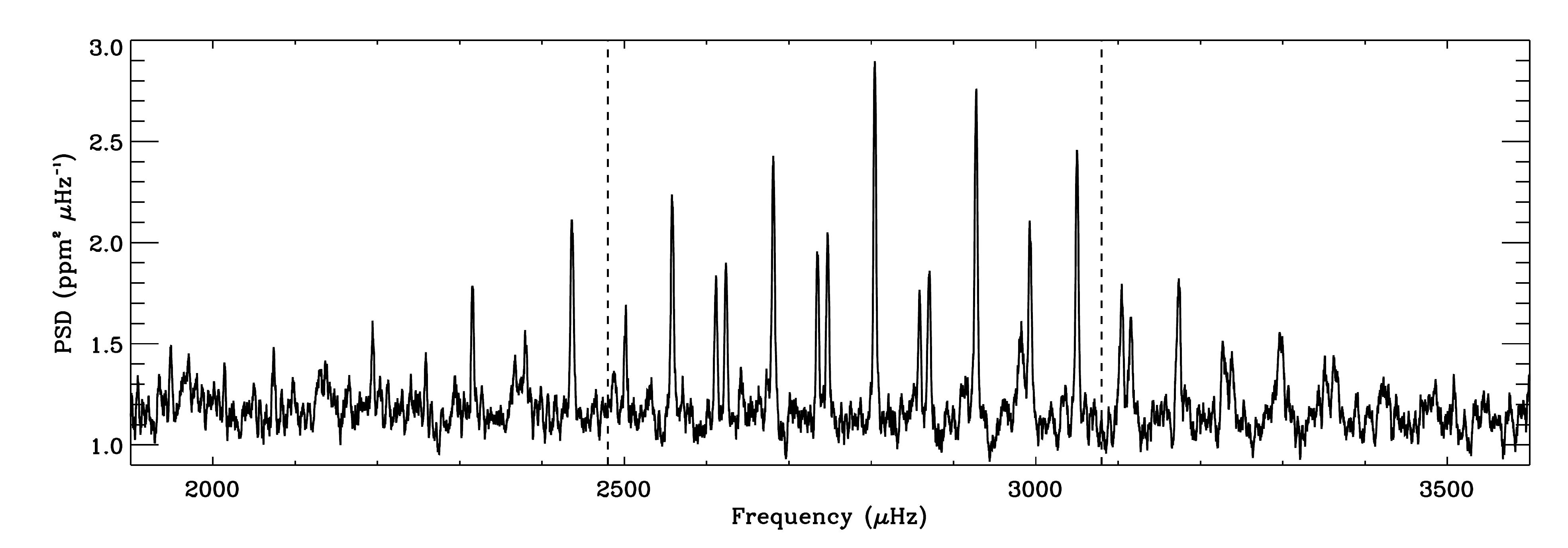}
\end{center}
\caption{\label{fig:fig1}
Power spectrum of KIC\,10644253 observed by the {\it Kepler} satellite centered around the frequency range of the measured acoustic oscillations and obtained by averaging the analyzed 180-day power spectra having a duty cycle larger than 80\% and smoothed over 4\,$\mu$Hz. The vertical dashed lines indicate the frequency range  chosen for the analysis of the frequency shifts.}
\end{figure*} 
% -------------------

% -------------------
\begin{figure*}[tbp]
\begin{center} 
\includegraphics[width=0.95\textwidth,angle=0]{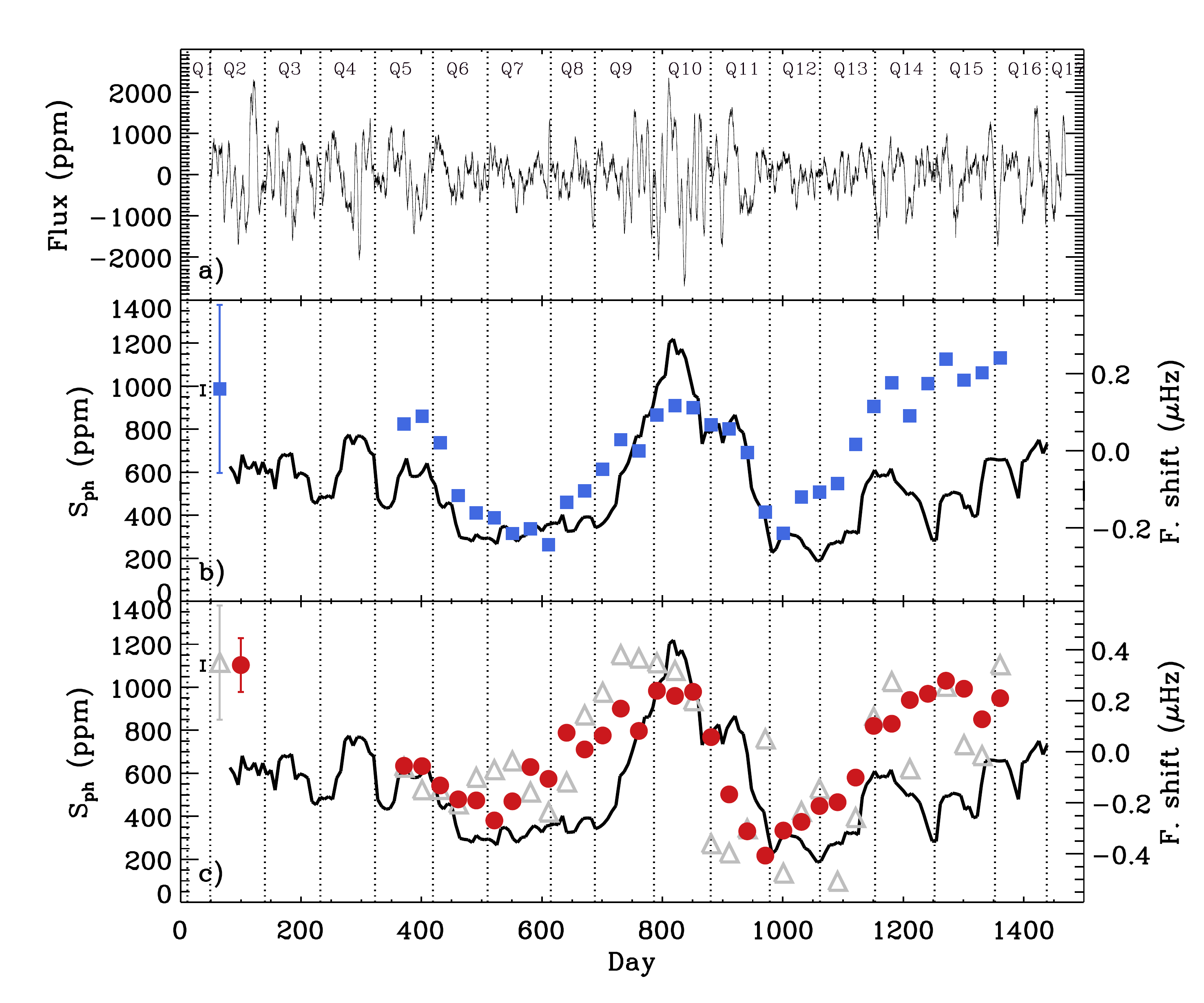}
\end{center}
\caption{\label{fig:fig2} 
(Top panel) Q2 to Q17 photometric long-cadence observations (in ppm) of KIC\,10644253 collected over 1411 days by the {\it Kepler} satellite as a function of time. 
(Middle panel) Magnetic activity photometric proxy $S_{\text{ph}}$ (in ppm, black line) of KIC\,10644253 estimated from the long-cadence observations as a function of time compared to the global frequency shifts of the visible modes obtained from the cross-correlation analysis (Method\,\#1, in $\mu$Hz, blue squares). The frequency shifts were extracted from the continuous short-cadence observations from Q5 to Q17. 
(Bottom panel) Same as the middle panel but for the frequency shifts of the individual angular degrees $l=0$ (grey triangles) and $l=1$ (red circles) extracted from the local peak-fitting analysis (Method\,\#2). The associated mean uncertainties are illustrated in the upper left-hand corner of each panel using the same color/symbol code. In the three panels, the vertical dotted lines represent the observational length of each {\it Kepler} quarter from Q0 to Q17.}
\end{figure*} 
% -------------------

%------------------------------------------------------------------------------------------------------------------------------
\section{Measurement of the magnetic activity variability}
\label{sec:global}
\subsection{Photometric magnetic activity proxy}
\label{sec:sph}
The surface rotation period of stars can be measured from photometric observations due to the signature of the passage of spots inducing periodic modulations in the luminosity. Moreover, 
as shown by \citet{garcia10} with the CoRoT observations of the F-star HD\,49933, the variability of the light curve, estimated by the standard deviation of the measurements and associated to the presence of spots or magnetic features rotating on the surface of the star, can be used as a global index
 of stellar activity. However, as the variability in the light curves can have different origins and timescales (such as magnetic activity, convective motions, oscillations, stellar companion), the rotation period needs to be taken into account in calculating a magnetic activity index derived from the {\it Kepler} observations. In this way, most of the variability is only related to the magnetism (i.e. the spots) and not
to the other sources of variability. \citet{mathur14b} showed that by measuring the light curve fluctuations over sub series of length $5\times P_\text{rot}$, where $P_\text{rot}$ is the rotation period of the star in days, provides a good proxy for studying the temporal evolution of magnetic activity. 
A photometric index of stellar magnetic variability ($S_\mathrm{ph}$) can be thus derived from the {\it Kepler} observations. Salabert et al. (in preparation) showed as well that such photometric proxy can be directly applied to the solar observations, like the ones collected by the Variability of Solar Irrandiance and Gravity Oscillations \citep[VIRGO;][]{frohlich95} photometric instrument onboard the {\it Solar and Heliospheric Observatory} \citep[SoHO;][]{domingo95} spacecraft. 
First of all, we used the entire duration of the {\it Kepler} observations up to Q17 to check that the surface rotation period of KIC\,10644253 is consistent with the rotation of $10.91 \pm 0.87$~days reported by \citet{garcia14a} using data up to Q14. The photometric activity proxy $S_\mathrm{ph}$ of KIC\,10644253 was then calculated over sub series of 54.55\,days, with a eight-time overlap of 6.82\,days, using 1411\,days of the continuous LC {\it Kepler} data from Q2 to Q17.  

%------------------------------------------------------------------------------------------------------------------------------
\subsection{Extraction of the oscillation frequency shifts}
\label{sec:fshift}
In order to study the temporal variations of the low-degree p-mode oscillation frequencies, the SC {\it Kepler} dataset was split into contiguous 180-day-long sub series, overlapped by 30 days.
We recall that as no observations of KIC\,10644253 were collected during the quarters Q2, Q3, and Q4, we started the analysis from Q5 up to Q17. The quarter Q1 was disregarded as it is only 33-day long. A total of 34 non-independent 180-day sub series of a frequency resolution of about 0.06 $\mu$Hz were thus analyzed with a mean duty cycle of 95\%.

\subsubsection{Cross-correlation analysis (Method\,\#1)}
\label{sec:cross}
The cross-correlation method returns a global value of the frequency shift for all the visible modes in the power spectrum (Method\,\#1).
First introduced by \citet{palle89} to measure the solar mean frequency shifts of the $l=0,1$, and 2 acoustic modes along the solar cycle from one single ground-based instrument, the cross-correlation method has since been widely used to study the temporal variation of the solar p modes from helioseismic observations \citep[see][and references therein]{chaplin07a}. More recently, the cross-correlation analysis was one of the methods employed to extract for the first time the frequency shifts of solar-like oscillations observed in a star other than the Sun, the F-type star HD~49933 observed by the CoRoT satellite \citep{garcia10}.

The power spectrum of each 180-day sub series\footnote{The {\it Kepler} SC time series was first transformed into a regular temporal grid.} was cross-correlated with a reference spectrum taken as the averaged spectrum of the independent spectra. The cross correlation were computed over 5 consecutive radial orders between 2480\,$\mu$Hz and 3080\,$\mu$Hz centered around the frequency of the maximum of the p-mode power excess \citep{chaplin14}. To estimate the frequency shift  $\langle  \delta \nu\rangle_{\mathrm{Method}\,\#1}$, the cross-correlation function was fitted with a Gaussian profile, whose centroid provides a measurement of the mean $l=0,1,$ and 2 mode frequency shift over the considered frequency range. The associated 1$\sigma$ uncertainties were obtained through Monte Carlo simulations as follows:  the reference spectrum was shifted in frequency by the mean frequency shift obtained for each sub series; then 500 simulated power spectra were computed by multiplying the shifted reference spectrum by a random noise with a distribution following a $\chi^2$ with 2 degrees of freedom. For each sub spectra, the standard deviation of all the shifts obtained from the 500 simulated spectra with respect to the corresponding reference spectrum was adopted as the 1$\sigma$ error bar of the fit \citep{regulo15}.
%------------------------------------------------------------------------------------------------------------------------------

%------------------------------------------------------------------------------------------------------------------------------
%------------------------------------------------------------------------------------------------------------------------------
\subsubsection{Mode peak-fitting analysis (Methods\,\#2 and \#3)}
\label{sec:pkb}
As explained in Section\,\ref{sec:cross}, the cross-correlation method returns a global value of the frequency shift for all the visible $l=0,1$ and 2 modes in the power spectrum averaged over a given frequency range. The study of the frequency shifts estimated through the analysis of individual mode frequencies provide additional results using independent methods. To do so, the individual acoustic frequencies were measured using both a local (Method\,\#2) and a global (Method\,\#3) independent peak-fitting analysis as described below.

The power spectrum of each sub series of SC data was obtained in order to extract estimates of the p-mode oscillation parameters. The individual frequencies of the angular degrees $l=0$, 1, and 2 provided by \citet{appourchaux12} were taken as initial guesses. The procedure to extract the p-mode parameters was performed by fitting sequences of successive series of $l=0$, 1, and 2 modes using a maximum-likelihood estimator as in \citet{salabert11} for the CoRoT target HD\,49933 (Method\,\#2). The individual $l=0$, 1, and 2 modes were modeled using a single Lorentzian profile for each of the angular degrees $l$. Therefore, neither rotational splitting nor inclination angle were included in the fitting model.  The height ratios between the $l=0$, 1, and 2 modes were fixed to 1, 1.5, and 0.5 respectively as the ones typically expected for {\it Kepler} stars \citep{ballot11a}, and only one linewidth was fitted per radial order $n$. The natural logarithms of the mode height, linewidth, and background noise were varied resulting in normal distributions. The formal uncertainties in each parameter were then derived from the inverse Hessian matrix. The mode frequencies thus extracted were checked to be consistent within the 1$\sigma$ errors with \citet{appourchaux12}. Additionally, we analyzed the temporal variations of the mode amplitudes and linewidths as they are known to vary with magnetic activity in the Sun \citep[see e.g.,][]{chaplin00,salabert06}, but the results are not reliable due to a poor determination of these parameters at the level of precision of the data. 

Alternatively, we also used a global peak-fitting technique to derive the mode parameters (Method\,\#3). We used the procedure described in \citet{ballot11b} in which the power spectrum is modeled as the sum of a background noise and oscillation modes. The background noise was described by an Harvey profile and each mode of a given angular degree $l$ and radial order $n$ was modeled as multiplets of Lorentzian profiles parametrized by a central frequency, a height, a linewidth, a rotational splitting  (common to all modes), and an inclination angle. We considered modes of degree $l=0,1$ and 2 and 11 consecutive radial orders. To reduce the parameter space, and as in Method\,\#2, the height ratios were fixed to 1, 1,5, and 0.5 respectively and one linewidth was fitted per radial order. All the parameters were estimated by performing a maximum a-posteriori estimation \citep[e.g.,][]{gaulme09}. This procedure was individually applied to each 180-day power spectrum.

The mean temporal variations of the frequencies were defined as the differences between the frequencies observed at different dates and reference values of the corresponding modes. The set of reference frequencies was taken as the average over the entire observations. The frequency shifts  $\langle  \delta \nu\rangle_{\mathrm{Method}\,\#2}$ and  $\langle  \delta \nu\rangle_{\mathrm{Method}\,\#3}$ thus obtained were then averaged over 5 consecutive radial orders between 2480~$\mu$Hz and 3080~$\mu$Hz, i.e. the same frequency range as the one used for the cross-correlation analysis in Section~\ref{sec:cross}. The frequency range was currently defined by the lowest and highest modes which were accurately fitted in all the analyzed 180-day power spectra. As the cross-correlation method returns mean frequency shifts unweighted over the analyzed frequency range, we decided to use unweighted averages of the frequency shifts extracted from the individual frequencies. We also note that we did not use the frequencies of the $l = 2$ mode because of its lower signal-to-noise ratio. 

We tried as well to extract the temporal variation of the amplitude of the p-mode envelope from the 180-day power spectra in the same manner as it was done for  HD\,49933 \citep{garcia10} using the A2Z pipeline \citep{mathur10}. Indeed, the amplitudes of solar-like oscillations in stars were observed to be suppressed with magnetic activity \citep[e.g.,][]{mosser09,garcia10,chaplin11,bonanno14b}. However, the absolute values of the oscillation amplitudes in the {\it Kepler} observations are subject to intrinsic changes due to the rotation of the satellite by 90$^\circ$ every 3 months. The consequence of this procedure is that a given target is observed with a different aperture in each of the CCDs, which have as well their own sensitivity and sources of noise. For these reasons, we were unable to extract reliable temporal evolution of the amplitude of the modes.

%------------------------------------------
\section{Results}
\label{sec:results}
The power spectrum of KIC\,10644253 averaged over the analyzed 180-day {\it Kepler} power spectra and centered between 1900 and 3600\,$\mu$Hz around its maximum oscillation power is shown on Fig.~\,\ref{fig:fig1}. We recall here that the frequency resolution is about 0.06\,$\mu$Hz.  For illustrative purpose, the spectrum was smoothed over 4\,$\mu$Hz. The LC data of KIC\,10644253 over the entire duration of the {\it Kepler} mission of ~$\sim$\,4 years is represented on the top panel of Fig.~\ref{fig:fig2}, where the periodic modulation corresponds to the passage of spots. The middle panel shows the global frequency shifts of the visible low-degree modes  as a function of time obtained from the cross-correlation analysis (Method\,\#1) of the SC data using 180-day non-independent sub series. The frequency shifts are compared to the photometric activity proxy, $S_\text{ph}$, estimated from the LC data (see Section\,\ref{sec:sph}). The bottom panel of Fig.~\ref{fig:fig2} shows the frequency shifts of the individual angular degrees $l=0$ (triangles) and $l=1$ (circles) extracted from the local peak-fitting analysis (Method\,\#2) using the same data as in Method\,\#1, compared as well to the photometric activity proxy $S_\text{ph}$. We note that comparable results within the error bars are obtained from the global peak-fitting analysis (Method\,\#3), but for illustrative purpose, we only show the results from Method\,\#2.

The photometric activity proxy, $S_\mathrm{ph}$, shows temporal variations going from periods of low activity ($\sim$\,300\,ppm) to periods of higher activity ($\sim$\,1200\,ppm) with clear rising and declining phases and a mean value of  $508.0\,\pm\,11.2$\,ppm. For comparison, the mean value of the photometric activity level of the Sun estimated from the VIRGO observations is $S_\mathrm{ph,\textsc{\sun}}=172.58\,\pm\,0.43$\,ppm, with values of $S_\mathrm{ph,\textsc{Max}\,\sun}\,=\,255.37\pm0.62$\,ppm and $S_\mathrm{ph,\textsc{Min}\,\sun}\,=\,65.91\pm0.23$\,ppm respectively at maximum and minimum of the solar activity (Salabert et al., in preparation). The p-mode frequency shifts of KIC\,10644253 are also observed to vary with time with a comparable modulation with a minimum-to-maximum swing of about 0.5\,$\mu$Hz. The three independent methods return consistent variations of the frequency shifts within the precision of the measurements. The cross-correlation method (Method\,\#1) returns however smaller frequency shifts than the ones obtained from the  peak-fitting analysis (Methods\,\#2 and \#3). This can be explained by the fact that the frequency shifts extracted from the cross-correlation method (Method\,\#1) correspond to a mean shift over the entire analyzed frequency range, and not only of the individual mode frequencies as in Methods\,\#2 and \#3.
Moreover, our results show that the individual $l=0$ and $l=1$ frequency shifts show similar variations although the $l=0$ frequency shifts present a larger scatter, which is translated in larger errors. This is probably due to the interplay with the nearby lower signal-to-noise ratio $l=2$ modes with a different surface weight. 
Furthermore, the frequency shifts, which are related to the response of magnetic and structural changes inside the star, are observed to increase before the surface photometric activity proxy does. Indeed, the rising phase in the frequency shifts is observed to start about 100 days before, while both observables appear to be more in phase during the falling phase of the modulation. We note that in the case of the Sun, time delays of about 60--90 days between the low-degree frequency shifts and the surface activity proxies were observed as well \citep{salabert09,salabert13}.

%%%%%%%%%%%%%%%%%%%%%%%%%%%%%%%
\begin{table}[t]
\caption{Most significant period ($P_\text{max}$) in days in the Lomb-Scargle periodograms calculated from the photometric activity proxy ($S_\text{ph}$) and 
the frequency shifts, $\langle  \delta \nu\rangle$, extracted from the three independent methods. The associated False Alarm Probability (FAP) are also given. Note that independent points only were used.}
\centering         
\begin{tabular}{ccc}        
\hline\hline
Observable   & $P_\text{max}$ & FAP \\
   & (day) & \\
   \hline
 $S_\text{ph}$ & 582 & 0.05 \\
 $\langle  \delta \nu\rangle_{\mathrm{Method}\,\#1}$  & 514 & 0.52 \\
 $\langle  \delta \nu_{l=1}\rangle_{\mathrm{Method}\,\#2}$ & 514 & 0.43 \\
 $\langle  \delta \nu_{l=1}\rangle_{\mathrm{Method}\,\#3}$ & 514 & 0.64 \\
 $\langle  \delta \nu_{l=0}\rangle_{\mathrm{Method}\,\#2}$ & 600 & 0.68 \\
 $\langle  \delta \nu_{l=0}\rangle_{\mathrm{Method}\,\#3}$ & 600 & 0.67 \\
 \hline
\end{tabular}
\label{tab:lombS}
\end{table}
%%%%%%%%%%%%%%%%%%%%%%%%%%%%%%%

% -------------------
\begin{figure}[tbp]
\begin{center} 
\includegraphics[width=0.45\textwidth,angle=0]{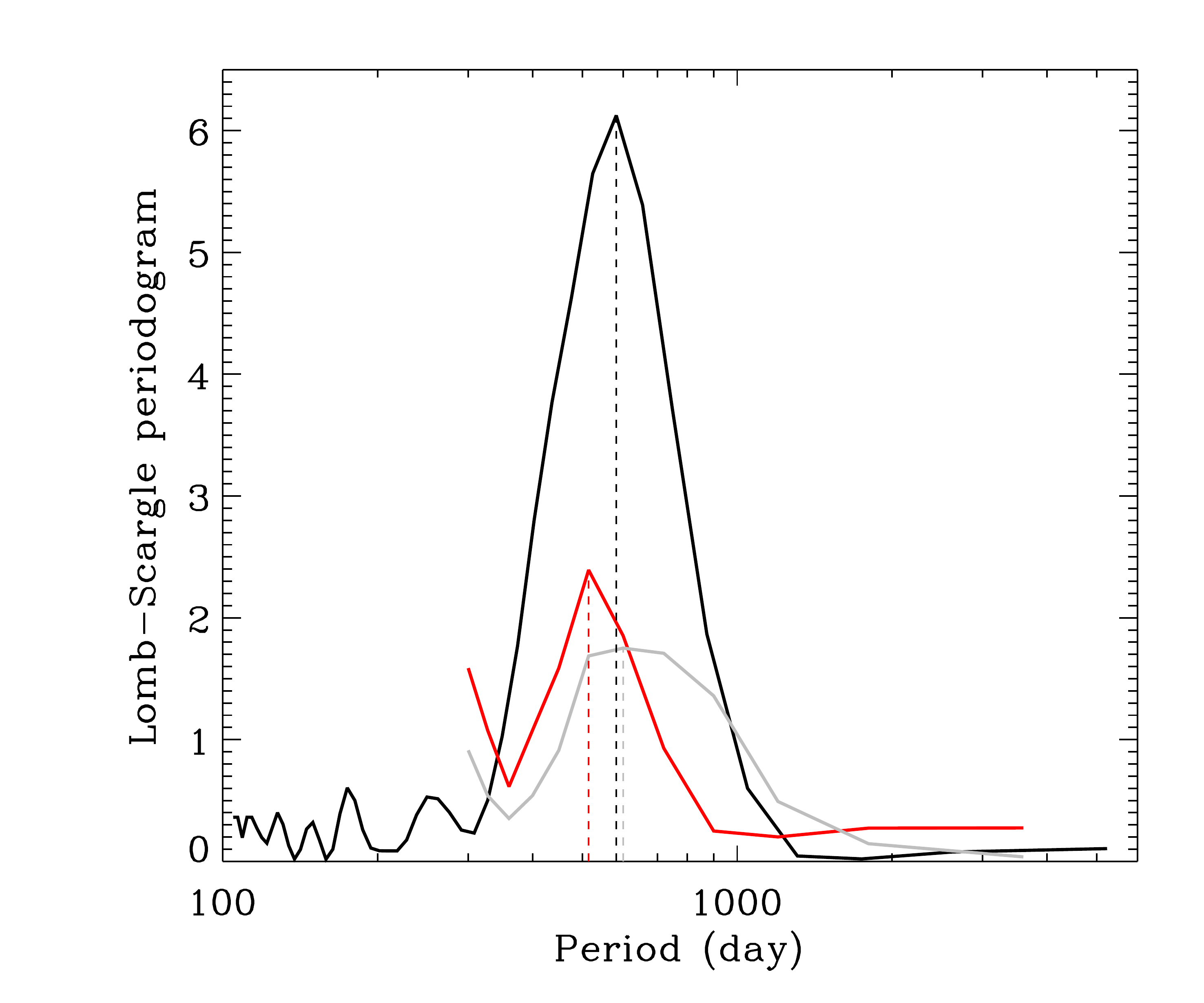}
\end{center}
\caption{\label{fig:fig3} 
Lomb-Scargle periodograms of the photometric activity proxy $S_\text{ph}$ (black) and of the frequency shifts of the individual $l=0$ (grey) and $l=1$ (red) modes extracted from the analysis of KIC\,10644253 using Method\,\#2. Note that independent points only were used. The vertical dashed lines correspond to the associated maximum of the periodograms.}
\end{figure} 
% ------------------- 

Lomb-Scargle (LS) periodograms were calculated from both the frequency shifts of the $l=0$ and $l=1$ modes and also from the photometric activity proxy using independent points only (Fig.\,\ref{fig:fig3}). We note that the LS periodograms were obtained by applying a four-time oversampling. The returned most significant periods and false alarm probabilities are given in Table\,\ref{tab:lombS}, including the ones obtained from the three methods used to extract the frequency shifts. The periodograms on Fig.\,\ref{fig:fig3} show a significant period centered around $\sim$\,550 days, i.e. $\sim$\,1.5\,years, given the resolution of the data. 
This period is comparable to the mean short-period variations of $\sim$\,1.7\,years found by \citet{egeland15} in the young 1\,Gyr-old G1.5V solar analog HD\,30495, which has a similar surface rotation period of 11.3 days and a long-period activity cycle of 12 years. They used a windowed Lomb-Scargle analysis to determine both quasi-periodic and intermittent variations over the 48-year period of the MWO observations, with the period of the variations ranging from $\sim$\,1.2 to $\sim$\,2.5\,years. However, as explained by \citet{egeland15}, the short-period variation in HD\,30495 was somewhat difficult to pin down. By comparison with KIC\,10644253, the $\sim$\,1.5-year variation also appears to be intermittent in the photometric activity proxy $S_\mathrm{ph}$, with one strong oscillation centered around day 850 (Fig.~\ref{fig:fig2}). Furthermore, this object has very comparable stellar properties to KIC\,10644253, including the surface rotation period. In addition, values reported by \citet{egeland15} are remarkably close to the two mean variations observed in the Sun's activity:  the 11-year cycle  and the quasi-biennial oscillation (QBO) 
 \citep[see e.g.][and references therein]{bazi15}. Moreover the solar QBOs observed in various activity proxies are also characterized by intermittence in periodicity and with amplitudes that vary with time, being larger around times of solar activity maxima. A tempting and simple hypothesis one can thus speculate about the solar analog KIC\,10644253 is that the $\sim$\,1.5-year variation observed here in the photometric activity proxy and the oscillation frequency shifts is actually the signature of the short-period modulation, or QBO, of its magnetic activity as it is observed in the Sun \citep[see e.g.,][]{fletcher10,broomhall12,simoniello13} and the 1-Gyr-old solar analog HD\,30495 \citep{egeland15}. We note however that in the case of the Sun, the frequencies shifts of the low-degree p modes between minimum and maximum of the 11-year solar cycle vary by about 0.4\,$\mu$Hz and by about 0.1\,$\mu$Hz over the QBO period. Longer observations are still needed in order to confirm the analogy of the activity temporal variations measured in KIC\,10644253 with the ones observed by \citet{egeland15} in HD\,30495 and in the Sun \citep{bazi15}.

%%%%%%%%%%%%%%%%%%%%%%%%%%%%%%%
\begin{table}[t]
\caption{Mean $l=1$ frequency shifts of KIC\,10644253 over two frequency ranges obtained from the individual frequencies (Method\,\#2) extracted between two periods of high and low activity.}
\centering         
\begin{tabular}{ccc}        
\hline\hline
 $\langle \delta\nu \rangle_{\mathrm{Method}\,\#2}$  &   \multicolumn{2}{c}{Frequency range} \\
  & \multicolumn{2}{c}{($\mu$Hz)} \\
  &   [2350--2700] & [2700--3100]  \\    
  \hline
 $l=1$ & 0.10$\pm$0.31 &  0.34$\pm$0.22\\     
\hline
\end{tabular}
\label{tab:mean_fshift}
\end{table}%
%%%%%%%%%%%%%%%%%%%%%%%%%%%%%%%

% -------------------
\begin{figure}[tbp]
\begin{center} 
\includegraphics[width=0.45\textwidth,angle=0]{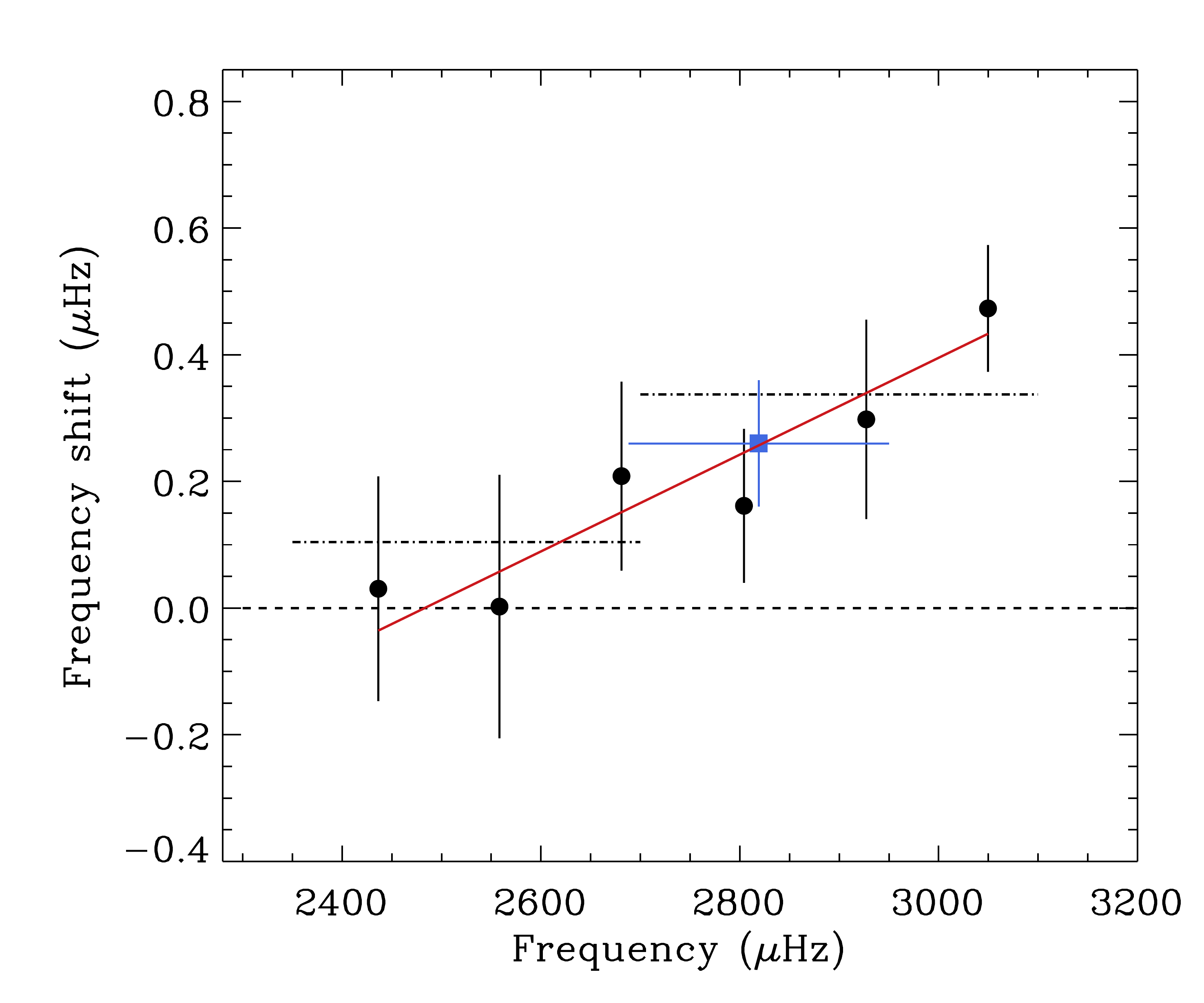}
\end{center}
\caption{\label{fig:fig4} 
Frequency shifts of the individual $l=1$ frequencies (black dots) of KIC\,10644253 as a function of frequency. The dot-dashed lines correspond to the weighted averages over two different frequency ranges represented by the total length of the lines, while the red solid line corresponds to a weighted linear fit. The blue square corresponds to the modeled value of the frequency shift at $\nu_\mathrm{max}$ (see Section~\ref{sec:discussion} and Table~\ref{tab:properties}).}
\end{figure} 
% -------------------

 %-----------------------
\begin{figure*}[t!]
\begin{center}
\includegraphics[width=1\textwidth]{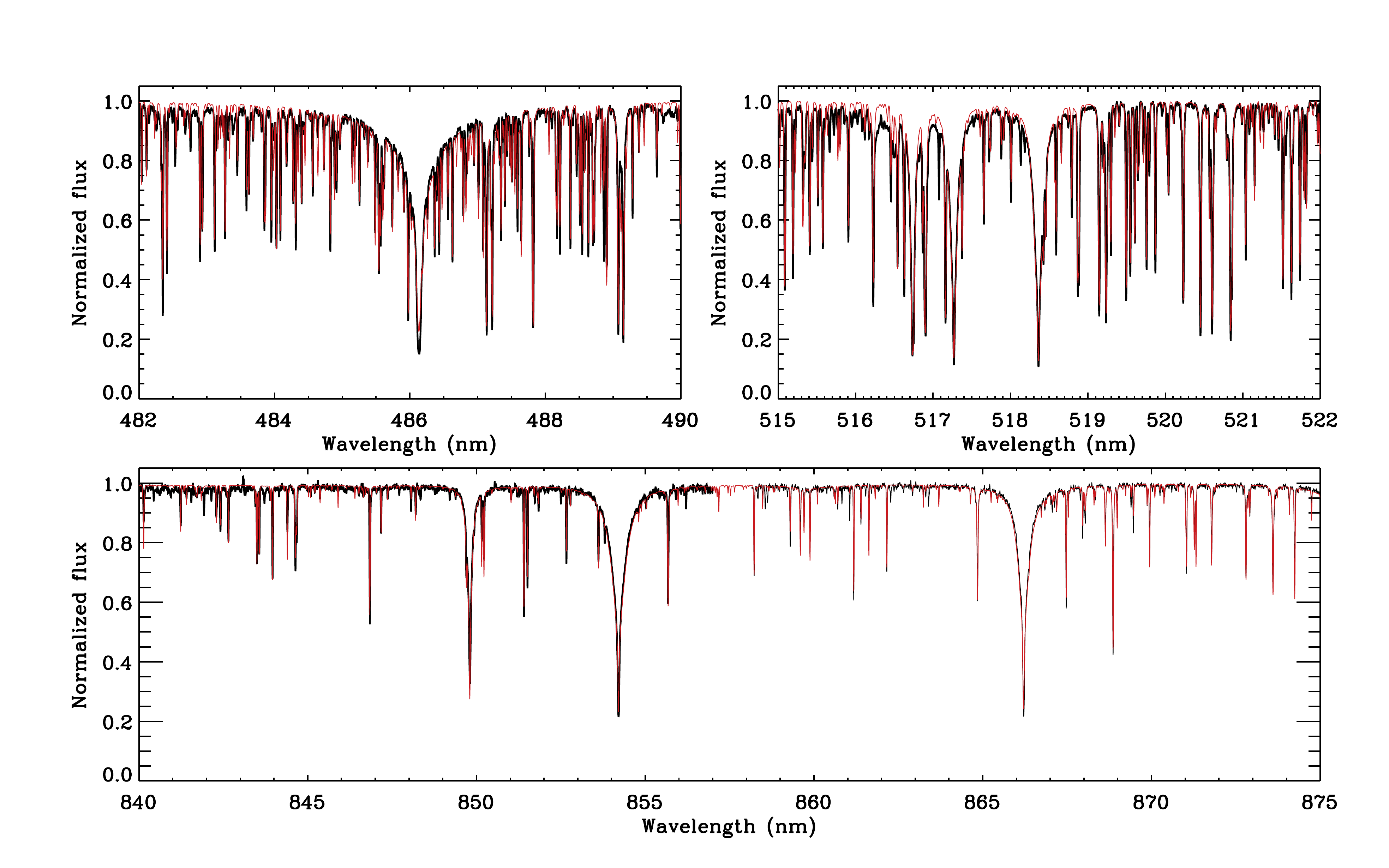}
\caption{Examples of the normalized spectrum of KIC\,10644253 (in black) collected by the \hermes spectrograph over three different  wavelength ranges, respectively around the lines H$_\beta$ (top-left panel), \ion{Mg}{ii} (top-right panel), and the Ca\,IRT triplet (bottom panel). The associated best fitting synthetic spectra are represented in red.
\label{fig:synthObsSpec}   }
\end{center}
\end{figure*}
%-----------------------

%-----------------------
\begin{figure*}[t!]
\begin{center}
\includegraphics[width=\textwidth]{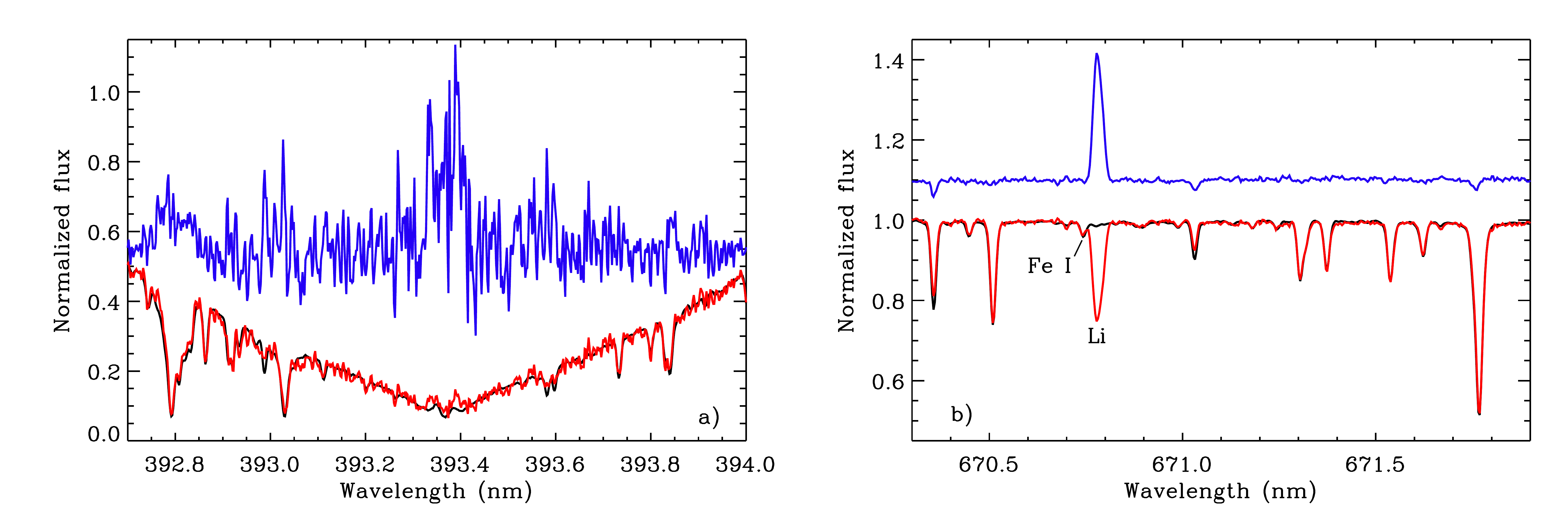} 
\caption{Comparison of the solar spectrum (black) to the spectrum of the solar analog KIC\,10644253 observed with the \Hermes spectrograph. The regions around the Ca\,K line and the lithium multiplet at 670.78\,nm are respectively shown in the left and right panels. The flux ratio (blue) of the two spectra between KIC\,10644253 and the Sun (left), and between the Sun and KIC\,10644253 (right) are also shown after shifting them by $-$0.45 and $+$0.1 respectively.
\label{fig:hermes_Mowgli} }
\end{center}
\end{figure*}
%-----------------------

The frequency shifts of the solar p modes are also observed to follow an inverse mode-mass scaling \citep{libbrecht90}, which is explained to arise from changes in the outer inner layers, with larger variations observed in the high-frequency modes than in the low-frequency modes  \citep[e.g.,][]{libbrecht90,chaplin98,salabert04}. Such frequency dependence was only observed today in one star other the Sun, the F-star HD\,49933 \citep{salabert11}. \citet{broomhall12} reported as well a frequency dependence of the solar QBO, although being weaker than the frequency dependence along the 11-year solar cycle. From the Fig.~5 in \citet{broomhall12}, we can estimate that the ratio is roughly of about a factor 3. In order to detect any possible frequency dependence in the frequency shifts observed here for the solar analog KIC\,10644253, oscillation mode frequencies were extracted over two periods of lower and higher activity in the same manner as described in Section~\ref{sec:pkb} (Method\,\#2) for six consecutive radial orders between 2350 and 3080\,$\mu$Hz. The frequency shifts at each radial order were thus obtained by simply subtracting the frequencies measured during high activity by the corresponding frequencies at low activity. We considered the $l=1$ modes only, because over the entired epoch considered here, the $l=1$ frequencies present the most significant swing between high and low activity. The results are shown on Fig.~\ref{fig:fig4}, where we see that the modes at higher frequency are more shifted than the ones at low frequency, reaching a maximum of about 0.5\,$\mu$Hz for the highest radial order analyzed here. The mean frequency shifts over two different frequency ranges and weighted by using the formal errors are also represented
and given in Table\,\ref{tab:mean_fshift}. Such frequency dependence closely resembles what is already observed for the Sun and the F-star HD\,49933, i.e., higher shifts at higher frequencies.

Furthermore, there are strong evidence that fast rotating young active stars show long-lived spots compared to the rotation period. As described in \citet{lanza14}, the decay of the auto-correlation of the light curve can provide indication on the lifetime of the active regions in a model-independent way. We found that  the spot pattern in KIC\,10644253 is observed to live around $\sim$\,300~days, so a lifetime much longer than the surface rotation of 10.9~days. Moreover, regarding the potentiality to study possible evidence of differential rotation, \citet{aigrain15} demonstrated with a blind hare-and-hounds exercise that all the existing methods developed today to extract the differential rotation in stars are not yet fully reliable. For these reasons, such analysis was not performed here, as further developments will be required to validate such methods.

%------------------------------------------------------------------------------------------------------------------------------
\section{Spectroscopic analysis}
The combined spectrum of KIC\,10644253 has an extremely high signal-to-noise ratio (SNR\,$\sim$\,280), allowing us to perform a detailed spectroscopic analysis. In Fig.~\ref{fig:synthObsSpec}, the observed spectrum is shown around the H$_\beta$ line at 486\,nm, as well as the lines \ion{Mg}{ii} 518\,nm and Ca-triplet in the near infrared around 850\,nm. These lines are particularly sensitive to the fundamental parameters. The lines needed for the elaborated spectral analysis of the lithium content as well as the chromospheric activity as discussed below are shown in Fig.\,\ref{fig:hermes_Mowgli} and compared to the solar spectrum.\\

\subsection{Chromospheric activity}
\label{sec:chromo}
The stellar activity level can be quantified through the \Sindex \citep[][and references therein]{duncan91}, which measures the strength of the emission in the cores of the Ca\,\textsc{ii}\,H and K lines in the near UV. 
The left panel of Fig.\,\ref{fig:hermes_Mowgli} compares the central region of the Ca\,K line in KIC\,10644253 to the solar spectrum obtained with \Hermes in April 2015 \citep{Beck15}. The instrumental factor to scale the \Sindex derived with \Hermes into the MWO system is $\alpha = 23\pm2$ \citep{Beck15}. It was estimated using observations collected with the \Hermes spectrograph of G2-type stars from the Mount Wilson Observatory archive \citep[MWO,][]{duncan91}.

The \Sindex was estimated from the unnormalized but order-merged individual spectra and converted into the MWO system. For each spectrum, the corresponding value is reported in Table\,\ref{tab:rv}. The associated mean \sindex\footnote{We note that the spectra \#\,3, 4, 5, and 6 were disregarded when estimating the mean value of the \sindex as they were obtained during poor weather conditions (see Table~\ref{tab:rv}).} is $\mathcal{S} = 0.213 \pm 0.008$ (Table\,\ref{tab:spectro}). Over the 180-day time base of the \hermes monitoring and given the few observed data points, the \sindex can be considered to remain stable within the $1\sigma$ dispersion of the measurements quoted as the error on the mean value. Comparing this value to the activity level of the Sun during its active phase of cycle~24, $\mathcal{S}_\sun=0.180$ \citep{Beck15} and represented in Fig.\,\ref{fig:hermes_Mowgli}, we can estimate that KIC\,10644253 is at least 18\% more active than the Sun. This is in agreement with what can be expected for young active stars with observed p-mode oscillations, while such young stars with no pulsations can be much more active than KIC\,10644253 \citep[e.g.,][]{frasca11,frohlich12}. However, since we have no indication on the phase of the activity modulation the \Hermes observations were taken, this number could actually be even larger at time of maximum of activity. It is also worth mentioning that the high-resolution spectropolarimetric BCool survey \citep{marsden14} measures surface magnetic fields and chromospheric activity proxies of cool solar-type stars. They showed that magnetic fields are likely to be detected for stars with a \sindex greater than about 0.2, which is the case of KIC\,10644253. 

The Mount Wilson \sindex is however affected by line blanketing in the continuum regions near the H and K lines, which biases the value for less massive stars.  To correct for this, another dimensionless ratio $R'_{\rm HK}$ was proposed in \citet{noyes84}, which applies a correction factor for the line blanketing and attempts to remove the photospheric contribution to the line core using an empirically derived relationship. The $\log R'_{\rm HK}$ is frequently used throughout the literature, and is more appropriate for comparing the chromospheric activity of KIC\,10644253 to the Sun. Using the mean $S = 0.213$ and the $(B-V)=0.59$ (see Table~\ref{tab:properties}) with the \citet{noyes84} relationships, we obtained $\log R'_{\rm HK} = -4.720\pm0.045$. The uncertainty in $\log R'_{\rm HK}$ was derived using standard error propagation methods. This again shows that KIC\,10644253 to be more active than the Sun, whose  $\log R'_{\rm HK}$ value ranges from $-4.86$ to $-4.95$ from solar maximum to minimum \citep{Hall07}.

%-----------------------
\begin{figure}[t!]
\begin{center}
\includegraphics[width=0.49\textwidth,angle=0]{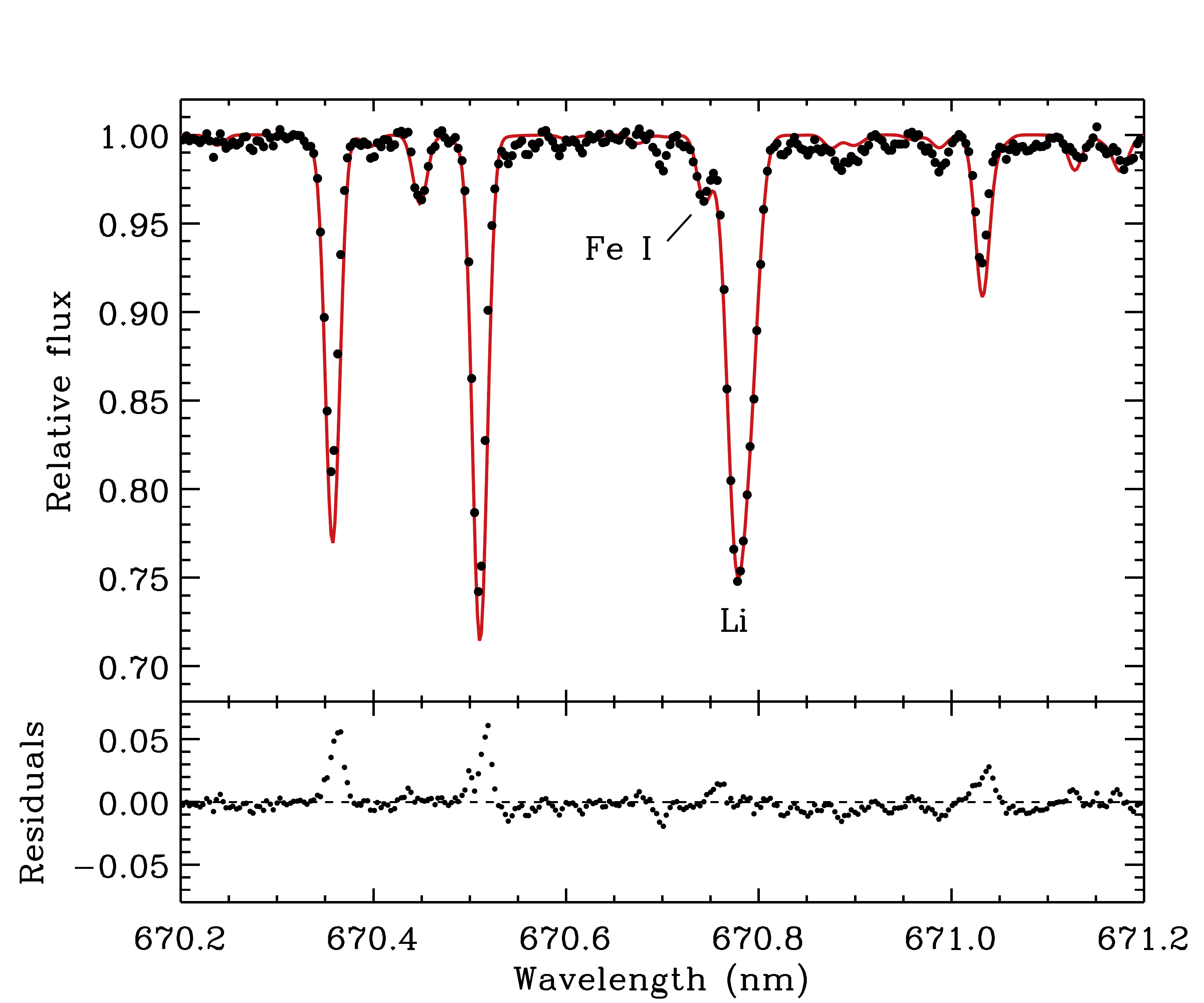}
\caption{(Top panel) Lithium doublet of the solar analog KIC\,10644253 observed with the \Hermes instrument (dots) around 670.7\,nm and associated spectral synthesis (red solid line) calculated using the spectroscopic set of atmospheric parameters given in Table~\ref{tab:spectro}. (Bottom panel) Residuals between the observed and the synthetic spectra.
\label{fig:liAbundance} }
\end{center}
\end{figure}
%-----------------------

\subsection{Fundamental parameters and lithium abundance}
To determine the fundamental parameters of KIC\,10644253, we performed a grid search by comparing the observed spectrum to a library of synthetic spectra, which were computed using the {\sc SynthV} radiative transfer code \citep{tsymbal96} based on the pre-computed grid of atmosphere models from the {\sc LLmodels} code \citep{shulyak04}. The search was based on a 40\,nm-wide wavelength range, containing the H$_\beta$ and the Mg-triplet lines as well as a $\sim$\,35\,nm-wide range containing the Ca infrared triplet at 860\,nm, using the \textit{Grid Search in Stellar Parameters} (GSSP\footnote{The GSSP package is available for download at https://fys.kuleuven.be/ster/meetings/binary-2015/gssp-software-package.}) software package \citep[][]{lehmann11,tkachenko12, tkachenko15}. The estimated effective temperature ($T_\mathrm{eff}=6006\pm100$\,K) and the surface gravity (log~$g=4.3\pm0.1$\,dex) are in excellent agreement with the results from \citet{bruntt12}. However, the metallicity reported by \citet{bruntt12} of $\mathrm{[Fe/H]}=0.12\pm0.03$\,dex differs from the value of $0.2\pm0.1$\,dex found here, but the difference falls within the quoted error of 0.1\,dex. The summary of the results obtained from the analysis of the \hermes observations is given  in Table~\ref{tab:spectro}, while the comparison between the observed and synthetic spectra is shown in Fig.\,\ref{fig:synthObsSpec}. 

Following \citet{jose13}, the lithium abundance of KIC\,10644253 was derived from the Li\,I resonance transition at 670.7\,nm by producing and comparing a synthetic spectrum to the observed \hermes spectrum, for the set of atmospheric parameters given in Table~\ref{tab:spectro}, to match a single-line equivalent width. This procedure requires an atmosphere model and physical parameters of a list of absorption lines to compute atomic chemical  abundances. The atmosphere models were interpolated in the Kurucz grid \citep{kurucz93} and the synthetic spectra were calculated with the MOOG routine by \citet{sneden73}.  We note that the default solar abundances used here were taken from \citet{anders89}. 

The lithium abundance\footnote{In this work, we used the standard definitions: [X/Y] = log(N$_\mathrm{X}$/N$_\mathrm{Y}$) - log(N$_\mathrm{X}$/N$_\mathrm{Y}$)$_\sun$, and A$_\mathrm{X}$= log (N$_\mathrm{X}$/N$_\mathrm{H}$)+12, where N$_\mathrm{X}$ is the number density of element X in the stellar photosphere.} of KIC\,10644253 was derived to be $\mathrm{A(Li) = log\,N(Li)}=2.74 \pm0.03$\,dex, which is in agreement with the value of \citet{bruntt12} of $\rm{A(Li)}-\rm{A(Li)}_\odot=1.74$, for a solar abundance $\rm{A(Li)}_\odot=0.96$ \citep{ghezzi09}.
The detailed modeling of the resonance doublet, composed of the lines of $^7$Li and $^6$Li at 670.7754 and 670.7766\,nm, respectively  is shown in Fig.\,\ref{fig:liAbundance}. 
The main sources of error on the Li abundance  are related to the uncertainties on the stellar parameters and equivalent-width measurement, however  $T_\mathrm{eff}$ is by far the dominant source of error. The lithium abundance of  2.74\,dex and a rotation period around 10.9~days are consistent with a young star, as stars in the Praesepe and Hyades clusters as determined by \citet{soderblom93}.  The two clusters are believed to be younger than 0.6\,Gyr-old. The rotation period is also consistent with the one obtained for 1.1 solar-mass stars of the 1 Gyr-old cluster NGC 6811 as described by \citet{meibom11}. This result suggests that KIC\,10644253 is a young object based on its Li depletion history \citep{ramirez12}, although the higher effective temperature can also contribute.

%-----------------------
\begin{table}[t!]
\caption{Summary of the results obtained from the analysis of the \Hermes spectroscopic observations of KIC\,10644253.}
\centering
\begin{tabular}{ll}
\hline\hline
Time base (days) 		& 180\\
Mean radial velocity	(km s$^{-1}$) & $-19.00\pm0.04$ \\
$T_\mathrm{eff}$  (K) & $6006\pm100$\\
log $g$ (dex) & $4.3\pm0.1$\\
$[\text{M/H}]$ (dex)  & $0.2\pm0.1$\\
$\upsilon\sin i$ (km s$^{-1}$) & $1\pm2$\\
A(Li) (dex) & $2.74\pm0.03$\\
\hline
\Ssymbol		   	& $0.213\pm0.008$\\
log R$_\mathrm{HK}'$ & $-4.720\pm0.045$ \\
\hline
\end{tabular}
\label{tab:spectro}
\end{table}
%-----------------------

%------------------------------------------------------------------------------------------------------------------------------
\section{Discussion}
\label{sec:discussion}
The mechanisms responsible for the frequency shifts of the acoustic oscillations along the solar magnetic activity cycle are still not completely clear. As shown by \citet{libbrecht90} for p modes at intermediate angular degree, the $l$-dependence of the shifts can be largely eliminated by considering changes normalized with the mode inertia, $I_{nl} \times\delta \nu_{nl}$. This indicates that the direct mechanism responsible for the shifts is located close to the solar surface. Moreover, they showed that the frequency dependence of the shifts was flattened once it was normalized by the mode inertia as well.
A decrease of less than 2\% in the radial component of the turbulent velocity in the outer layers is assumed to explain the frequency increase of the low- and intermediate-degree acoustic modes with solar magnetic activity \citep{kuhn00,dziem01,dziem05}.

\begin{figure}[t!]
\begin{center}
\includegraphics[width=0.49\textwidth]{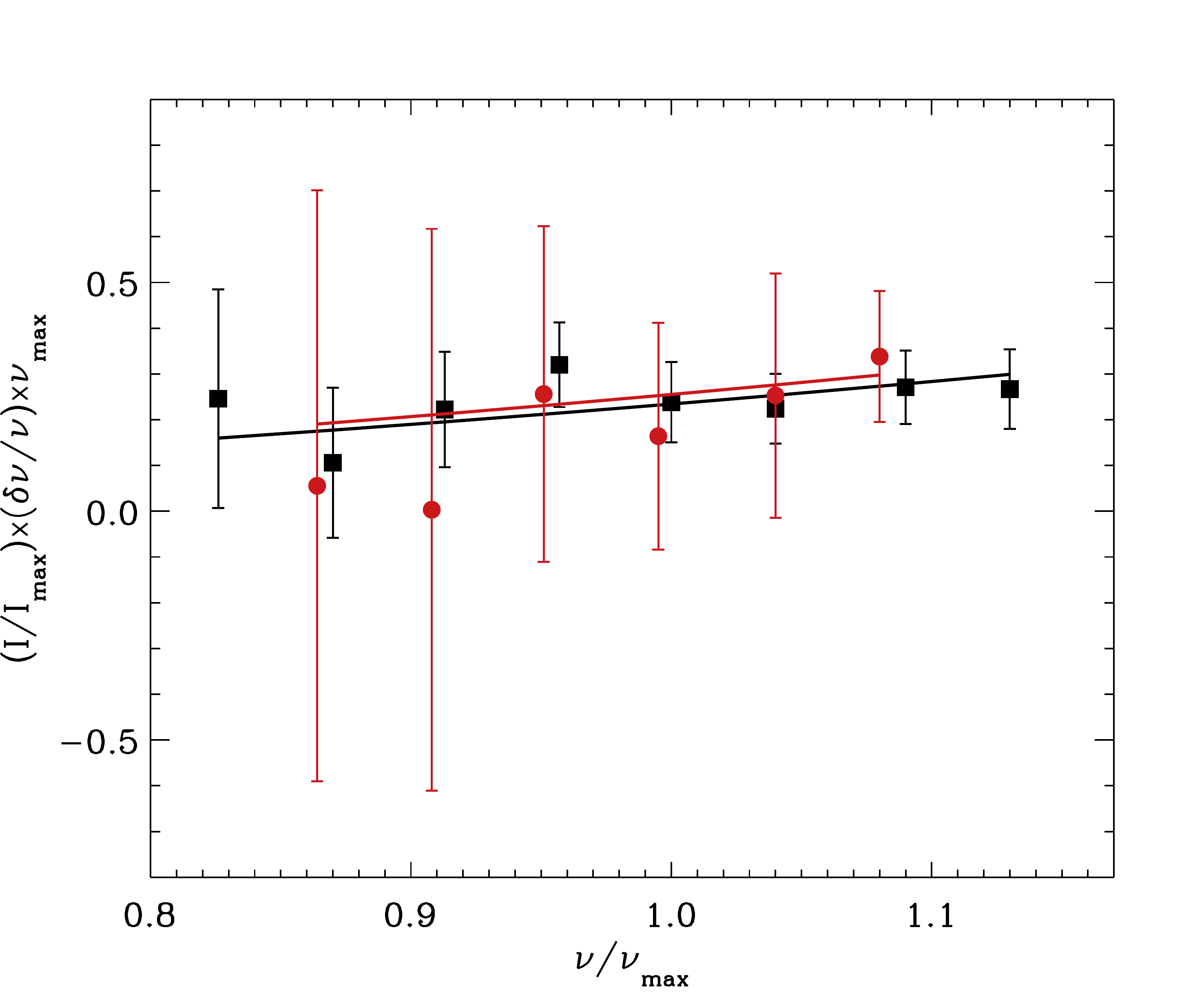}
\caption{Dimensionless normalized $l=1$ frequency shifts $(I_{nl}/I_\mathrm{max}) \times \delta \nu_{nl} /\nu_{nl} \times  \nu_\mathrm{max}$ as a function of $\nu/\nu_\mathrm{max}$ for both the Sun (black squares) and KIC\,10644253 (red dots). The solid lines correspond to the associated weighted linear regressions of the form $A_0 \nu^2$.
\label{fig:inertia} }
\end{center}
\end{figure}

The dimensionless normalized frequency shifts defined as $(I_{nl}/I_\mathrm{max}) \times \delta \nu_{nl} /\nu_{nl} \times  \nu_\mathrm{max}$ are shown on Fig.~\ref{fig:inertia} as a function of $\nu/\nu_\mathrm{max}$ for both the Sun \citep{salabert15} and KIC\,10644253. The frequency $\nu_\mathrm{max}$ at the maximum oscillation power was used as it is linearly related to the acoustic cut-off frequency \citep{belkacem11}.
We only considered here the $l=1$ modes as explained in Section~\ref{sec:results}. The solid lines correspond to weighted linear regressions of the form $A_0\nu^2$, where $A_0$ is
the fitted value of the frequency shift at $\nu_\mathrm{max}$. The frequency dependence was assumed to follow the simple prescription by \citet{gough90}. We found $A_0= 0.23$\,$\pm$\,0.03\,$\mu$Hz and $A_0=0.26$\,$\pm$\,0.10\,$\mu$Hz for the Sun and KIC\,10644253 respectively. The $A_0$ value for KIC\,10644253 at $\nu_\mathrm{max}$ is compared  to the observed frequency dependence of the $l=1$ modes shown on Fig.~\ref{fig:fig4}.
The measured frequency shifts for the Sun and the solar analog KIC\,10644253 have a comparable frequency dependence as seen on Fig.~\ref{fig:inertia}. However, the theoretical predictions, such as the ones proposed by \citet{chaplin07b} and \citet{metcalfe07} suggest a slightly higher value for KIC\,10644253 than for the Sun. But given the approximations considered in their theoretical estimations and the uncertainties on the extracted frequency shifts, both observations and predictions are consistent.

The frequency shifts measured in this work are clearly associated to the $\sim$\,1.5-year
magnetic modulation observed in the photometric activity proxy $S_\mathrm{ph}$. When compared to the Sun, the physical
mechanism for these shifts could be related either to a long-period variation over a decade or so, or to a short-period variation such as the QBO.
This question is relevant since as already mentioned, \citet{egeland15} reported dual modulations of about 1.7 and 12 years in the young solar analog HD\,30495, which is a very similar object to KIC\,10644253 in many ways.
Moreover, the differences measured in the frequency shifts of the Sun along the 11-year cycle and the QBO can be attributed to differences in the structure changes induced by the magnetic activity \citep{broomhall12,salabert15}. 
Indeed, the degree and frequency dependence of the solar frequency shifts were observed to be different for the QBO and the 11-year cycle. The larger frequency shifts of the QBO measured for the quadripolar $l=2$ modes \citep{broomhall12} were explained to be associated with an equatorial and deeper concentration of the structural changes in the QBO, as also suggested by \citet{ulrich13} and \citet{mcintosh15}.  
Furthermore, KIC\,10644253 is likely to have a very inclined rotation axis (see Table~\ref{tab:properties}), although we cannot be more precise as the differences between the measurements exceed their reported uncertainties. The projected velocity $\upsilon \sin i$ of 1\,km s$^{-1}$\ measured here with \Hermes (Table~\ref{tab:spectro}) returns an angle of about $11^{\circ}$, confirming nonetheless the high inclination angle of this star in relation of the line of sight.
The contribution of the zonal modes will be then smaller and hence a weaker degree dependence than for the Sun is expected for KIC\,10644253.
Moreover, based on \citet{broomhall12}, the dimensionless frequency shifts of the solar QBO calculated as in Fig.~\ref{fig:inertia} suggests that their frequency dependence is negligeable. The slope of the frequency shifts measured in KIC\,10644253 are similar to the one observed along the solar 11-year cycle of the Sun, although the
large errors prevent to exclude a flatter slope and hence a deeper location of the source as in the QBO.
We note as well that in consequence of the highly-inclined rotation axis of KIC\,10644253, the activity proxy $S_\mathrm{ph}$ derived here should be considered as a lower limit of its photospheric activity.

%-----------------------
%-----------------------

%------------------------------------------------------------------------------------------------------------------------------
\section{Conclusions}
We analyzed the photometric {\it Kepler} observations of the young (1\,Gyr-old) solar analog KIC\,10644253, collected in both long and short cadences. We studied its temporal magnetic variability by estimating different proxies of activity. The light curve was directly analyzed to estimate the photometric activity through the measure of the global proxy $S_\mathrm{ph}$ over sub series of $5\times P_\mathrm{rot} = 54.55$\,days. In addition, the temporal variability of the oscillation frequencies of the low-degree acoustic modes was obtained through the analysis of consecutive 180-day power spectra using peak-fitting and cross-correlation independent methods, and the associated frequency shifts $\langle \delta\nu \rangle$ extracted. We showed that both the photometric activity proxy $S_\mathrm{ph}$ and the frequency shifts $\langle \delta\nu \rangle$ present the signature of magnetic activity with a significant temporal variability. Furthermore, a modulation of about 1.5\,years over the duration of the {\it Kepler} observations was measured in both observables of about 900\,ppm for $S_\mathrm{ph}$ and 0.5\,$\mu$Hz for the frequency shifts. This 1.5-year variation could actually be the signature of the short-period modulation, or quasi-biennal oscillation, of its magnetic activity as it is observed in the Sun \citep[see e.g.,][]{fletcher10} and the 1-Gyr-old solar analog HD\,30495 with very close stellar properties \citep{egeland15}.
Moreover, the comparison between the magnitude and the frequency dependence of the
frequency shifts measured for KIC\,10644253 with the ones obtained for the Sun indicates that the same physical mechanisms are involved in the sub-surface inner layers in both stars. 

In addition, complementary high-resolution spectroscopic observations of KIC\,10644253 were collected by the \Hermes instrument mounted to the 1.2\,m \textsc{Mercator} telescope located in La Palma (Canary Islands, Spain). The analysis of the emission in the cores of the Ca\textsc{ii}\,H\&K lines showed that KIC\,10644253 is about 18\% chromospherically more active than the mean emission of the Sun with a \sindex calibrated into the Mount Wilson Observatory system of $0.213\pm0.008$, thus confirming that it is an active object. Moreover, the high lithium abundance of $2.74\pm0.03$\,dex and the effective temperature of $6006\pm100$\,K  of the solar analog KIC\,10644253 mean that the lithium at the surface has not been depleted yet by internal processes \citep{ramirez12}.  This is thus validating its young age  ($\sim$\,1\,Gyr-old) estimated from seismology and in agreement with a rotation period of about 11 days from gyrochronology \citep{meibom11}.

Finally, the young solar analog KIC\,10644253 has been put on a long-term monitoring program with the \Hermes spectrograph, which will provide measurements of the chromospheric \sindex over a long period of time. Moreover, new photometric observations in the near future with the Transiting Exoplanet Survey Satellite (TESS) instrument \citep{ricker15} scheduled to be launched around 2017--2018 will again allow a monitoring of the photospheric activity of this star and to also extract additional measurements of the acoustic oscillation frequency shifts. That will provide a longer temporal coverage of KIC\,10644253's activity allowing us to compare it in greater detail with the Sun and HD\,30495.

%------------------------------------------------------------------------------------------------------------------------------
\begin{acknowledgements}
The authors wish to thank the entire {\it Kepler} team, without whom these results would not be possible. Funding for this Discovery mission is provided by NASA's Science Mission Directorate. 
 We also thank all funding councils and agencies that have supported the activities of KASC Working Group\,1, as well as the International Space Science Institute (ISSI). The ground-based observations are based on spectroscopy made with the Mercator Telescope, operated on the island of La Palma by the Flemish Community, at the Spanish Observatorio del Roque de los Muchachos of the Instituto de Astrof\'isica de Canarias. 
The research leading to these results has received funding from the European Community's Seventh Framework Programme ([FP7/2007-2013]) under grant agreement no. 312844 (SPACEINN) and under grant agreement  no.  269194  (IRSES/ASK). The research leading to these results has also been supported by grant AYA2012-39346-C02-02 of the Spanish Secretary of State for R\&D\&i (MINECO). DS and RAG acknowledge the financial support from the CNES GOLF and PLATO grants. PGB acknowledges the ANR (Agence Nationale de la Recherche, France) program IDEE (n$\degr$ ANR-12-BS05-0008) "Interaction Des Etoiles et des Exoplan\`etes". JDNJr acknowledges support from CNPq PQ 308830/2012-1 CNPq PDE Harvard grant.
RE is supported by the Newkirk Fellowship at the High Altitude Observatory.
SM acknowledges support from the NASA grant NNX12AE17G.
DS acknowledges the Observatoire de la C\^ote d'Azur for support during his stays. This  research  has  made  use  of  the
SIMBAD database, operated at CDS, Strasbourg, France.
\end{acknowledgements}

%-------------------------------------------------------------------

\end{document}